# Strongly Modulated Exfoliation and Functionalization of MXene with Rational Designed Groups in Polymer: A Theoretical Study

Qiye Guan[a], Hejin Yan[a], Yongqing Cai[a,*]

[a] Joint Key Laboratory of Ministry of Education, Institute of Applied Physics and Materials Engineering, University of Macau, Taipa, Macau, China;

**ABSTRACT**: As emerging atomically ultrathin metal compounds, MXenes show great promise for catalysts and nanoelectronics applications due to the abundant surface terminations and high metallic conductivity. However, the tendency of the interlayer adhesion and suffering from environmental disturbances significantly limit their endurance and efficiency. Herein via conducting first-principles calculations, we explore surface passivation and exfoliation of MXene via polymers which have been experimentally proven to promote the performance. Nine kinds of monomers together with the typical MXene $Ti_3C_2T_2$ (T= None, O, F, OH, $F_{0.5}O_{0.5}$) as prototype composites are explored with respect to the adsorption and charge transfer associated with energetics and chemical redox, respectively. Our work shows that naked $Ti_3C_2$ MXene has a strong ability to cleave and decompose the monomers. Surface functionalized $Ti_3C_2F_2$, $Ti_3C_2FO$, and $Ti_3C_2O_2$ have a weak binding with monomers through only van der Waals force, whereas $Ti_3C_2(OH)_2$ also exhibits strengthened binding for some monomers. Specific functional groups in the monomer, such as the halogen, sulfur, and hydroxyl groups or a relatively planar aromatic structure, largely contribute to the adsorption. We reveal that the functionalization through polymer would alter the carriers' density via interfacial charge transfer in MXenes. While the naked $Ti_3C_2$ and $Ti_3C_2(OH)_2$ donate electrons to the polymers, the $Ti_3C_2F_2$, $Ti_3C_2FO$, and $Ti_3C_2O_2$ receive small amounts of electrons transferred from the polymer, highly depending on the types of the monomers. The varying ability of charge transfer and exfoliation energy of different monomers implies great flexibility for designing polymers to exfoliate the MXene and modulate the carrier densities which is highly desired for altering conductivity, dielectric properties, and promoted endurance.

## INTRODUCTION

Since the successful exfoliation of single-layer graphene, tremendous progress and capability gained in recent years to isolate dispersed atomically thin layers enables the massive production of high-quality 2D materials like hexagonal boron nitride,[1] transition metal dichalcogenides[2] which in turn enables a vast range of possibilities in various fields, such as energy storage, catalysis, ion sieving membrane, and sensors, etc. As a typical 2D material, transition metal carbides and nitrides known as MXene have a universal formula $M_{n+1}X_nT_x$ (where M is an early transition metal, X is carbon or nitrogen, $n$ is 1-3, and T is surface termination of oxygen, hydroxyl, or fluorine). Generally, MXenes are produced by selectively etching from the parent compounds MAX, where A is an A-group element, such as Al or Si. MXenes have been proven to have great potential and application[3-5] in energy storage, electromagnetic interference (EMI) shielding, and catalysis because of their highly active surfaces, good conductivity, abundant functional groups, and facile synthesis[6,7]. Moreover, a large specific area and versatile functionalities make MXenes an excellent host for plenty of molecule adsorbates and other 2D materials additives, expanding the potential application field of MXene derivatives.

To remedy intrinsic limitations of MXene, such as easy agglomeration and poor stability in the environment of pure MXene, and simultaneously promote the performance such as the energy and power capacity in lithium-ion batteries, flexibility in soft sensors, and stability in high-temperature for electromagnetic interference (EMI) shielding, MXenes are usually functionalized with other materials like transitional metal dichalcogenides,[8] carbon nanotubes,[9] and polymers.[10-12] Among these composites, the hybridization of MXene with polymer has been identified as a practical strategy with low cost, simple fabrication, and adjustable functionalities.[13-18] Through intercalation of polymers which expands the interlayer spacing, adhesion of MXene flakes will be hindered.[19,20]

The introduction of functional polymer groups was also found to promote electronic, mechanical, and chemical performances as well. Due to the diversified properties of MXene combined with polymers, they have been already applied in various fields, like sensing, energy storage, sewage purification, catalysis, electromagnetic absorption, and shielding. Monomers that share the same functional groups with polymers will have similar effects on the target MXene. For energy applications, the addition of polymer would stabilize and intercalate MXene, which prevents stacking and collapse of MXene, and in the meantime, the reaction sites and adsorption ability of the composites will be increased. Liu et al. first reported the $Ti_3C_2T_x$/perfluorosulfonic acid (Nafion) composite as proton transfer membranes with improved conductivity.[21] $Ti_3C_2T_x$/polyvinyl alcohol (PVA) composite film shows significantly enhanced tensile strength compared to the pure $Ti_3C_2T_x$ or PVA films.[22] $Ti_3C_2T_x$/polyacrylamide (PAM) nanocomposite films exhibit a high conductivity up to $3.3 \times 10^{-2}$ S/m.[23] $Ti_3C_2T_x$/polyethylene terephthalate (PET) used for supercapacitors has good cycling stability with a retention of 87% capacitance after 1000 cycles.[24] PVA[22], polypyrrole (PPy)[25], or polyfluorene derivatives (PFDs)[26] combined with MXene are proved to be good supercapacitor electrode materials, especially for PFDs/MXene composites which improve the mass capacitance density to 150% comparing the bare MXene. For EMI shielding MXene/polymer, through combination with polymers like epoxy (EP)[27] or polyaniline (PANI)[28], the composite will be corrosion-resistant, lightweight, and will have enhanced electromagnetic shielding performance. Polytetrafluoroethylene (PTFE) /$Ti_3C_2T_x$/polyimide (PI) sandwich-like composite was also shown to be a promising EMI shielding material.[29] Polymethyl methacrylate (PMMA) /$Ti_3C_2T_x$/ZnO composites exhibit a superior dielectric constant of 435 with low energy leakage.[30] The combination of $Ti_3C_2T_x$ and conductive polymer (CP), including polyaniline (PANI), poly para-aminophenol (PPAP), polypyrrole (PPy), and polythiophene (PTh), allows a great enhancement of thermal conductivity, heat capacity and thermal diffusivity.[31] In the field of sensing, MXene/polymer composites also show promising applications in humidity, gas sensors, and wearable sensors. Composites that functionalized MXene with polyvinyl alcohol (PVA)/polyethyleneimine (PEI)[32] exhibit high sensitivity to toxic air pollutants due to the porous structure. $Ti_3C_2T_x$/polyvinyl butanal (PVB) fabricated for wearable sensors, exhibits a wide detecting window of voltage down to 0.1 mV.[33] $Ti_3C_2T_x$/polyacrylic acid (PAA) composites self-healing hydrogels show distinctive thermosensation-based actuation upon near-infrared illumination.[34]

It is evident that the hybridization of MXene with polymers is technically viable and experimentally intensively studied, but as far as we know, the theoretical mechanism underlying the functionalization of polymer is unknown and lacking albeit theoretical methods on layered interactions are satisfyingly mature.[35-40] Actually, most of the modulated macroscopic properties in polymer functionalized MXene have scenario on the atomic scale. Mechanisms behind the decreased conductivity in polymer-covered $Ti_3C_2T_x$ flakes compared with pure MXene,[34] and enhanced migration of ions[23] or proton transport[21] in surface-functionalized MXene films could be rooted in the charge transfer and degree of adhesion between the MXene and polymer layers. Similarly, for EMI shielding MXene/Polymer, the electronic conductivity and hydrophobic self-cleaning performance are also sensitive to the interaction between the MXene and functional groups of the polymers, such as halogen atoms, carboxyl group (-COOH), and hydroxyl group (-OH).[29] The dielectric property of MXene/Polymer is highly dependent on interfacial polarization between the polymer and MXene.[30] Therefore, for a clear explanation of the modulated behaviors upon the combination of MXene and polymers, an atomic-scale knowledge of the interaction between the MXene and polymer, especially the role of different functional groups in the polymer, is indispensable.

In this work, we aim to gain an atomic-scale view of those MXene/polymer hybrids based on first-principles calculations. We selected nine kinds of monomers including PAA, PAM, PCFE, PTFE, PANI, PI, PMMA, PVA, and PTh, which are chosen as representative of diverse functional groups in popular polymers. We are particularly interested in their abilities of adhesion and charge transfer to the MXene with prototype forms of $Ti_3C_2T_2$ (T=None, O, OH, F, or $F_{0.5}O_{0.5}$). We found that both the functional groups on the MXene (F, O, OH) and monomers play decisive roles. In addition, we also explore the possibility of using functional polymers to exfoliate the MXene by examining the energetics of monomers above MXene, as inspired by polymers being able to exfoliate $MoS_2$.[37] Through calculations of the theoretical exfoliation energy, dipole moment, density of states (DOS), and differential charge transfer (DCD) between two components, we disclose the role of polymers in $Ti_3C_2T_2$/Polymer composites.

**METHOD**

Spin-polarized first-principles calculations are carried out by the plane wave code Vienna ab initio simulation package (VASP)[41] with the projector augmented wave (PAW) method. Exchange-correlation energy is performed under the generalized gradient approximation (GGA) with the form of Perdew-Burke-Ernzerhof (PBE).[42] DFT-D3 functional with Grimme[43] correction is employed to describe the weak van der Waals (vdW) interaction between monomers and MXene. An energy cutoff of 450 eV and an energy convergence of $10^{-4}$ eV are used in all calculations. All the structures are optimized until the forces exerted on each atom are <0.005 eV Å$^{-1}$. A 3×3×1 supercell is adopted, and the Brillouin zone K-point mesh is set as 2 ×2 × 1 for structural optimization, then a 12 ×12 ×1 K-point mesh is applied for electronic structure computations. The thickness of the vacuum region is set to >15 Å to avoid interference of periodic images.

## RESULTS AND DISCUSSION

**Structural Properties and Adsorption Configurations.** To identify the effect of functional units in different polymers, we simulate the monomer which is the smallest replica representing a specific type of chain polymer. Starting with exploring the combination/adsorption behavior of MXene and monomers, we constructed nine different $Ti_3C_2T_2$/Monomer composites including monolayer and bilayer structures. All the isolated monomers representing different functional groups are optimized firstly as shown in **Figure 1**. For each case, we considered several different possible adsorbing configurations (**Figures S1-S5**) and chose the lowest energy configuration for analysis. We calculated the adsorption energy $E_{ads}$ which reflects the strength of interaction between the monomer and MXene:

$$E_{ads} = E_{total} - E_{mono} - E_{MXene} \tag{1}$$

where $E_{total}$ is the total energy of the constructed composites, $E_{mono}$ and $E_{MXene}$ are the monomer's and MXene's energy, respectively. As shown in **Table 1**, pristine $Ti_3C_2$ and functionalized $Ti_3C_2(OH)_2$ are more prone to bonding with monomers compared to F or O functionalized $Ti_3C_2T_2$. The difference of the $E_{ads}$ could be ascribed to the functional groups both on the MXene and monomers. For $Ti_3C_2O_2$, $Ti_3C_2F_2$, and $Ti_3C_2FO$, the O and F terminated surfaces lead to hindrance of interlayer combination between neighboring layers of few-layer MXene, while for OH, the abundant H atoms on the surface lead to the relatively strong van der Waals interaction with monomers, even covalent bonding with chlorotrifluoroethylene(CFE) accompanied with the broken of the C-Cl bond. The breaking of π bonds in acrylic acid(AA) and methyl methacrylate(MMA) induces a large adsorption energy of 3.3 ~ 3.4 eV. Similarly, the functional groups in monomers promote the adsorption with bare $Ti_3C_2$ upon hybridization. Most adsorption processes show a formation of bonds between the oxygen atom in monomers (from carboxyl group (acrylic acid(AA)), amide group (acrylamide(AM), succinimide(IM)), ester group (methyl methacrylate(MMA)), and hydroxyl group (vinyl alcohol(VA)) with surface Ti atom of $Ti_3C_2$. While for tetrafluoroethylene (TFE), chlorotrifluoroethylene(CFE), and thiophene(Th) the adsorption is stronger due to the more stable bond between the S or halogen (F, Cl) atoms with the Ti compared with oxygen-terminated monomers.

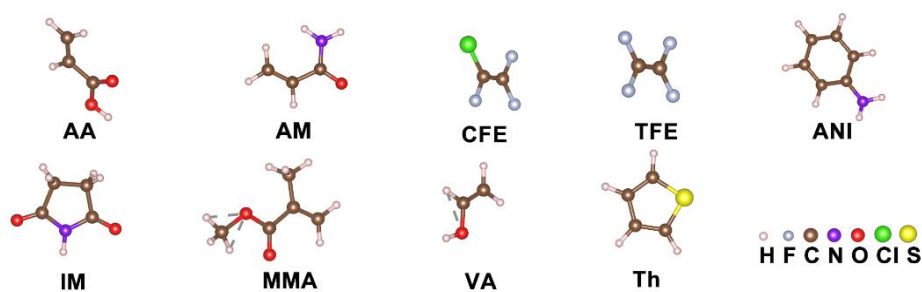

**Figure 1.** Schematic diagram of nine kinds of optimized monomers. Acrylic acid (AA), acrylamide (AM), chlorotrifluoroethylene (CFE), tetrafluoroethylene (TFE), aniline (ANI), succinimide (IM), methyl methacrylate (MMA), vinyl alcohol (VA), thiophene (Th).

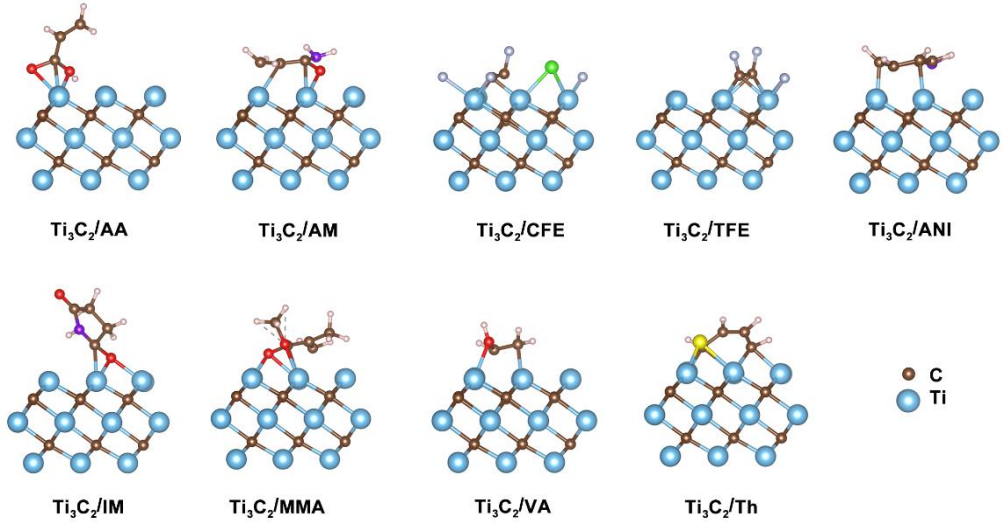

**Figure 2.** Structural atomic models for the adsorption of monomers above the surface of monolayer $Ti_3C_2$. Note the structural decomposition occurring for the CFE, TFE, and Th.

**Table 1.** Adsorption energy (eV) of polymer monomer above MXene monolayer and bilayer (marked by *).

|  | $Ti_3C_2$ | | $Ti_3C_2O_2$ | | $Ti_3C_2F_2$ | | $Ti_3C_2FO$ | | $Ti_3C_2(OH)_2$ | |
| --- | --- | --- | --- | --- | --- | --- | --- | --- | --- | --- |
| **AA** | -4.89 | -5.05* | -0.29 | -0.30* | -0.23 | -0.31* | -0.20 | -0.20* | -3.27 | -3.33* |
| **AM** | -4.34 | -4.11* | -0.53 | -0.58* | -0.46 | -0.48* | -0.52 | -0.51* | -1.80 | -1.89* |
| **CFE** | -12.86 | -12.72* | -0.39 | -0.40* | -0.34 | -0.35* | -0.34 | -0.33* | -4.30 | -4.32* |
| **TFE** | -9.53 | -9.70* | -0.32 | -0.32* | -0.28 | -0.29* | -0.28 | -0.27* | -0.34 | -0.40* |
| **ANI** | -3.94 | -4.09 | -0.88 | -0.90* | -0.62 | -0.63* | -0.73 | -0.71* | -1.07 | -1.09* |
| **IM** | -1.91 | -2.71* | -0.62 | -0.62* | -0.54 | -0.54* | -0.56 | -0.55* | -0.73 | -0.84* |
| **MMA** | -4.47 | -4.60* | -0.71 | -0.71* | -0.62 | -0.62* | -0.65 | -0.64* | -3.34 | -3.43* |
| **VA** | -2.16 | -2.44* | -0.46 | -0.46* | -0.28 | -0.29* | -0.32 | -0.34* | -0.78 | -0.65* |
| **Th** | -6.65 | -6.88* | -0.52 | -0.52* | -0.42 | -0.44* | -0.45 | -0.44* | -0.79 | -0.88* |

For $Ti_3C_2$, we initially considered the physisorption and put those molecules away from the surface, surprisingly all these nine molecules finally become chemically adsorbed on the $Ti_3C_2$ surface by forming covalent bonds as shown in **Figure 2** and **Figure S7**. Interestingly, we find the CFE and TFE monomers, become decomposed with the halogen atoms forming covalent bonds with exposed Ti atoms at MXene (Cl and F), indicating the strong destructive and cleaving ability of the pristine $Ti_3C_2$ MXene surface. Our work suggests that CFE, TFE, and Th monomers have a similar chemical cleavage tendency allowing the creation of -Cl, -F, and -S residual species in MXene. These anions strongly bound at the surface of MXene would alter the charge distribution and create dipole moment by forming polar Ti-X bonds (X=Cl, F, S) in the proximity of the surface. Unintentional introduction of these atomic residuals through mixing MXene with polymer would alter the electromagnetic response and account for the EMI shielding of MXene-polymer hybrid as masking or screening materials.[28,39]

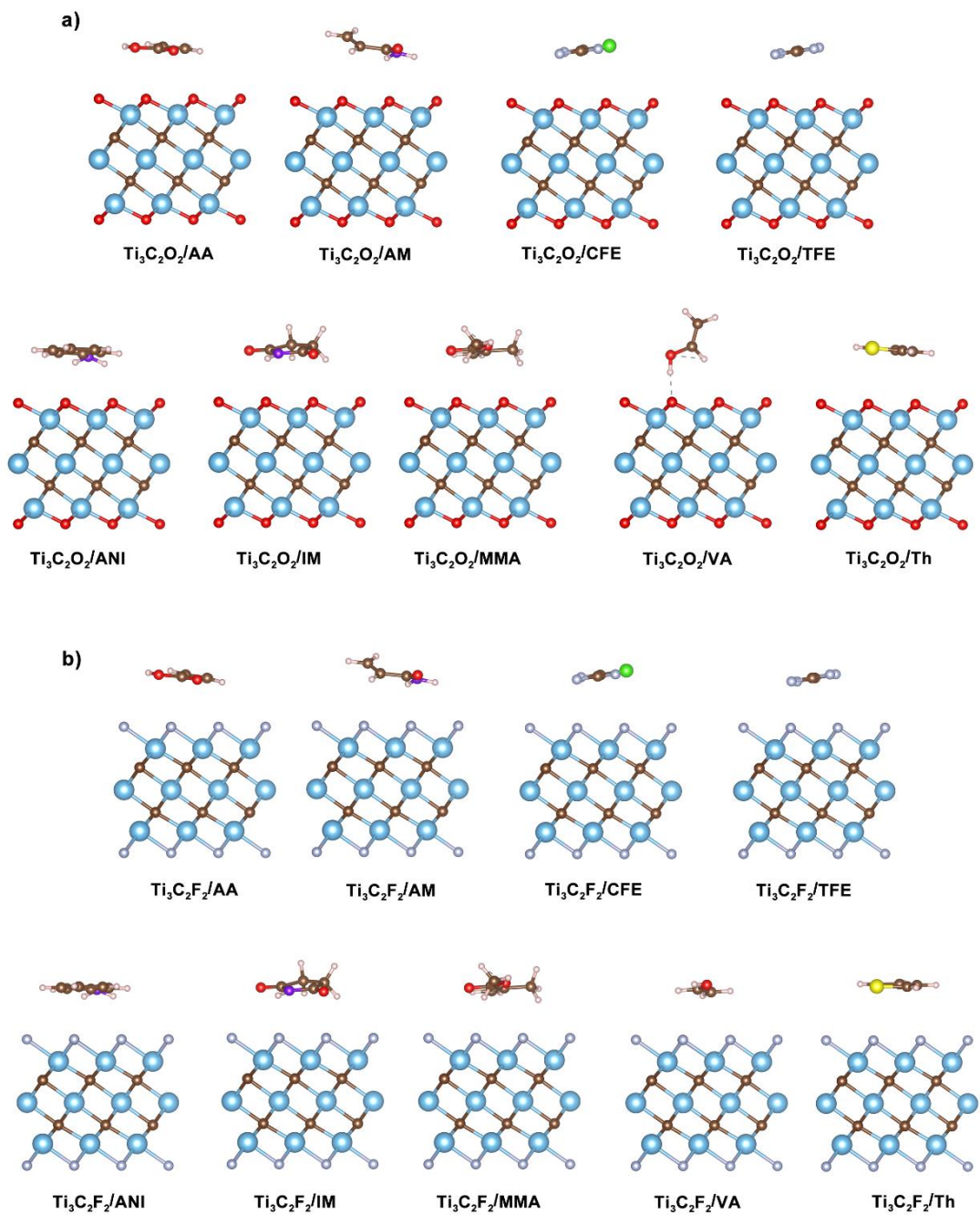

**Figure 3.** Structural models of a) **monolayer** Ti$_3$C$_2$O$_2$/Monomer and b) Ti$_3$C$_2$F$_2$/Monomer.

Experimental synthesized MXene through HF etching is usually surface functionalized by chemical residuals like oxygen, hydroxyl, and fluorine. Here, to simulate the effects of surface residuals on the adsorption of polymer groups, the $Ti_3C_2$ with oxygen ($Ti_3C_2O_2$), fluorine ($Ti_3C_2F_2$), oxygen-fluorine mixed ($Ti_3C_2FO$), and hydroxyl ($Ti_3C_2(OH)_2$) terminated MXenes are also investigated. As exhibited in **Figure 3** and **Figure S6**, the F-, O-, and FO- terminated composites all are weakly connected through vdW interaction. This is strikingly different compared to the pristine $Ti_3C_2$ where covalent bondings with polymer functional groups are dominant. In the meantime, the formation of hydrogen bonds is noticeable between monomers and -F (-O) atoms. For the O-terminated MXene, the polymer monomer tends to have similar adsorbing behaviors as compared to the F-terminated surface. In contrast, for OH-functionalized composite, the interaction between the monomer and OH groups is more significant, as shown in **Figure S6 b.** Even strong covalent bonding appears in AA and MMA, in which the $\pi$ bond associated with -COOR groups could be broken and reorganized with dissociated H atoms in -OH on the surface of the MXene. This behavior also appears in CFE with the breaking of C-Cl bond and forming of C-H bond. Similar adsorption behaviors are detected for all bilayer MXene/monomer composites upon the uptake of the polymer monomers(**Figures S7-S11**).

**Table 2.** Exfoliation energy (meV/Å$^2$) of $Ti_3C_2T_2$/Monomer composites.

|      | $Ti_3C_2$ | $Ti_3C_2O_2$ | $Ti_3C_2F_2$ | $Ti_3C_2FO$ | $Ti_3C_2(OH)_2$ |
|------|--------|----------|----------|---------|------------|
| **Bare** | 229.48 | 32.99 | 26.19 | 28.94 | 45.29 |
| **AA**   | 171.73 | 29.83 | 24.62 | 26.67 | - |
| **AM**   | 180.04 | 27.03 | 21.02 | 22.98 | 26.76 |
| **CFE**  | 73.03  | 28.70 | 22.39 | 25.15 | - |
| **TFE**  | 114.67 | 29.52 | 23.10 | 25.86 | 48.01 |
| **ANI**  | 183.24 | 22.78 | 19.13 | 20.54 | 37.72 |
| **IM**   | 206.95 | 26.05 | 20.00 | 22.44 | 40.55 |
| **MMA**  | 176.84 | 24.85 | 18.90 | 21.43 | - |
| **VA**   | 204.43 | 27.40 | 22.97 | 25.21 | 40.70 |
| **Th**   | 149.09 | 27.29 | 21.50 | 23.89 | 40.25 |

**Energetics and Exfoliation Behavior with the Uptake of Polymer Monomer.** Owing to its exposed Ti atoms at the surface and metallic nature, pristine $Ti_3C_2$ MXene layers have a high tendency to agglomerate. In applications such as electrocatalysis and supercapacitor, such interlayer adhesion would reduce the number of active sites and should be prevented. To increase the surface volume ratio, efficient exfoliation of layers of MXene thus is highly desired. Organic long-chain polymers are ideal candidates for exfoliation owing to their structural flexibility and selective modulated functional groups. Organic proteins have been demonstrated for exfoliating 2D $MoS_2$.[37] Here we show the possibility of the exfoliation of MXene with polymers by examining the energetics (exfoliation energy). To simplify the model and analysis, we use the monomers which are accurate enough to capture the different effects of functional groups in longer chain polymers.

The formation of MXene/Monomer composites induced by the monomers is closely related to the exfoliation energy shown in **Table 2**. The energy density of exfoliation is calculated by the following equation:[44,45,46,47]

$$E_{exf} = (E_{II} + E_{III} - E_I)/A \tag{2}$$

where $E_{II}$, $E_{III}$, and $E_I$ are total energies of the exfoliated, the remaining, and the original structure (shown in **Figure S12**). $A$ is the basal area. As given in **Table 2**, pristine $Ti_3C_2$ undoubtedly is the most favorable for fabricating MXene/Polymer composite with the strongest exfoliation energy for all kinds of monomers compared with surface functionalized ones. As a benchmark of exfoliation ability, we also calculated the exfoliation energy without monomers for all five types of 2-layer $Ti_3C_2T_2$. Noticed that the experimentally measured exfoliation energy of graphite[48] is 22.25~23.72 meV/Å$^2$, and the theoretically calculated exfoliation energy of 2D materials like $MoS_2$[49] is 28.7~33.5 meV/Å$^2$, which is comparable to the surface functionalized MXenes. Compared with the pristine $Ti_3C_2$ which has higher exfoliation energy of 229.42 meV/Å$^2$, the OH, F, or O functionalized $Ti_3C_2$ with 26 ~ 45 meV/Å$^2$ is much easier to be exfoliated.

In detail, with the lower exfoliation energies of 73.03 and 114.67 meV/Å$^2$, $Ti_3C_2$ is easily exfoliated through the interaction with CFE and TFE. This ensures the high probability of the formation of a layer-by-layer structure during synthesis, which enables the performance of EMI shielding material by the more effective conversion of electrical energy to Joule heat, and the more stable electrothermal deformation under the applied voltage.[25] While for $Ti_3C_2$/AA, $Ti_3C_2$/AM, $Ti_3C_2$/ANI, $Ti_3C_2$/IM, and $Ti_3C_2$/VA composites, the exfoliation energies are much higher around 170 ~ 210 meV/Å$^2$. This increase will directly weaken the exfoliation of $Ti_3C_2$ and leads to the dispersion of the polymer chains around the $Ti_3C_2$ flakes. As shown in **Figure 2**, the adsorption of monomers above $Ti_3C_2$ involves the formation of strong covalent bonds which means it is hard to completely remove monomers from the exfoliated $Ti_3C_2$/Monomer hybrid.

For surface functionalized composites, the density of exfoliation energy is overall much lower than the $Ti_3C_2$. Owing to the weak vdW interaction, the exfoliated MXene monolayer would largely maintain its structural integrity. For O-, or FO- terminated composites, ANI distinguishes out with the lowest exfoliation energy. With the planar phenyl group attached to the amino group, the lone pair on the nitrogen is partially delocalized into the π system of the benzene ring which results in negatively charged C atoms. This form of charge separation thus results in the enhanced interaction between the C atom of ANI and O atoms on the surface of $Ti_3C_2O_2$ or $Ti_3C_2FO$, and accordingly lower exfoliation energy. Interestingly, different from the O- or F- terminated MXene/Monomer composites for which the weak vdW interaction dominated the combination, OH-terminated $Ti_3C_2$ tends to be more active during the binding process. The bond broken of C-Cl in CFE and π bond broken in -COOR of AA and MMA are detected, accompanied by dissociation of the surface H atom. This would result in the inaccuracy of the calculation of original bilayer MXene, thus for these three types, the exfoliation energy is omitted. Due to the tight covalent bonding between these monomers and $Ti_3C_2(OH)_2$, similar to $Ti_3C_2$/Monomers, compact composites would be formed while still easy to be exfoliated because of the surface OH which weakens the layer combination. Our work predicts the dispersive MXene layers upon the introduction of polymers. Indeed, experimentally synthesized $Ti_3C_2T_x$/PVA[22] showed a weakened conductivity due to the presence of polymer chains between the $Ti_3C_2T_x$ flakes. Moreover, because of the well-dispersed structure, the Ti$_3$C$_2$T$_x$/PAM also showed the potential to facilitate the ions of the electrolyte to fast immerse and diffuse.[23]

To get a more precise view of how functional groups of monomers could influence the composite, we also calculated the dipole moment of all these nine molecules as shown in **Figure 4**. Our result shows that there is a non-monotonic dependence of the exfoliation energy with the dipole moment of the monomer. Interestingly, we find that monomers with a small dipole could also result in appreciable adsorption above Ti$_3$C$_2$. This is not surprising as the adsorption of the polymer above the unpassivated surface is mainly covalent while the dipole moment is correlated with the weak dispersive force. CFE, which has the second smallest dipole moment around 0.67 D, is broken during the adsorption process, forming new bonds of Ti-F. Our result explains the experiment result which reported the existence of the F element on the surface of the MXene Ti$_3$C$_2$ in MXene/Polymer composite.[27] Interestingly, TFE, which has a large dipole moment, 3.57 D, also exhibits strengthened exfoliation behavior forming Ti-F and Ti-Cl bonds. This facile decomposition of CFE and TFE also indicates that halogen, especially the fluorine element, is suitable for coating of MXene, fabricating a more compact structure. Similarly, the sulfur atom of Th also departs from the molecule to form bonds with Ti at the MXene. The destructed thiophene molecule above Ti$_3$C$_2$ allows the formation of the new Ti-C bonds with the average bond length of 2.2 Å and the Ti-S bond with the 2.5 Å, which is strengthened compared with the origin π bond in the thiophene and covalent bonds between the S and C atoms.

We found when the dipole moment is beyond 1.95 D, the exfoliation energy is flattened with neglectable variations except for monomers containing halogen atoms.

Unanimous with exfoliation analysis shown in **Figure 2** and **Figure 3,** with a moderate interaction, charge redistribution weakens when the dipole moment of the monomer is larger than 2 D. In those molecules, the π electrons are more prone to be localized toward these functional groups (-CONH$_2$, -COOR, RCONHCOR',-OH and -NH$_2$). Therefore only the oxygen's lone pair electrons are involved in the surface bonding with Ti atoms and accordingly a redistribution of charge. While for monomers whose dipole moment is lower than 2 D, the adsorption process tends to be accompanied by the electron redistribution within the backbone of the monomers, like the breaking of old bonds and the formation of new bonds. Similarly, for bilayer MXene/Molecule modules, the same behavior is found as shown in **Figure S7**.

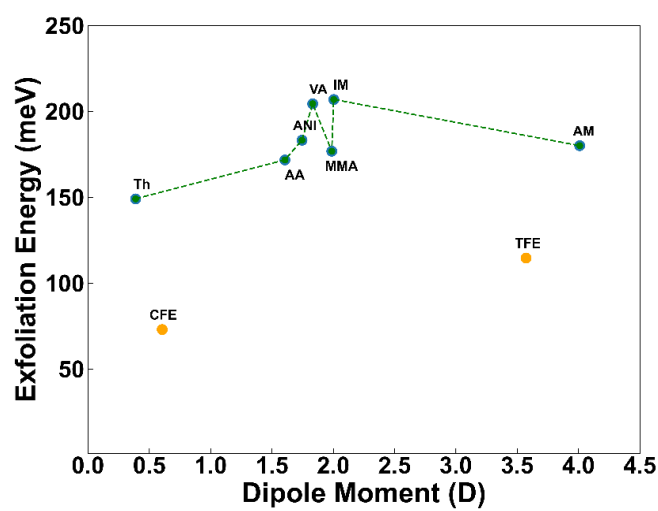

**Figure 4.** The exfoliation energy of monolayer Ti$_3$C$_2$/Momoner as a function of the dipole moment of monomers.

**Density of States and Charge Transfer.** We next examine the electronic properties of MXene upon the uptake of the polymer monomers. The charge transfer between those functional monomers and MXene will affect the carriers' density and conductivity of the host material. **Figure 5a** shows the local density of states (LDOS) of monolayer

MXene and AA (Other types could be found in **Figures S13-S17**). We notice that all the monomers have almost no contribution at the Fermi level owing to the isolated levels. The DOS of all the composites shows a high peak at the $E_f$, implying a good metallicity being maintained. However, for $Ti_3C_2F_2$, $Ti_3C_2FO$, $Ti_3C_2(OH)_2$, and $Ti_3C_2O_2$, the metallicity is weakened due to the passivation of F, O, or OH atoms. As expected, the monomers have a stronger effect on the LDOS of $Ti_3C_2$ than those of surface functionalized with levels being highly broadened (**Figure S13-S17**). Each monomer changes the shape of the LDOS of $Ti_3C_2$ around the Fermi level, implying a strong modulating capability. For a visualization of the change of metallicity after the decoration, the variation of the intensity of DOS at the Fermi level ($E_f$) is calculated and shown in **Figures 5b, 5c**. For F, or O functionalized $Ti_3C_2$, the change is negligible. Consistent with the adsorption behavior, for all these three types, ANI exhibits the largest DOS change at the $E_f$ among all the monomers. However, for OH-terminated $Ti_3C_2$, due to the higher reactivity, the DOS change at the $E_f$ is subtle, especially for CFE, AA, and MMA. Notably, for the AA, CFE, TFE, MMA, and Th monomers, there exist molecular levels close to the Fermi level, these close-lying states may form resonant channels for conduction and thermally activated electronic emission under finite voltage, inducing modulated conductivity. The DOS of bilayer MXene/Molecule is coinciding with the monolayer as shown in **Figures S18-S22.**

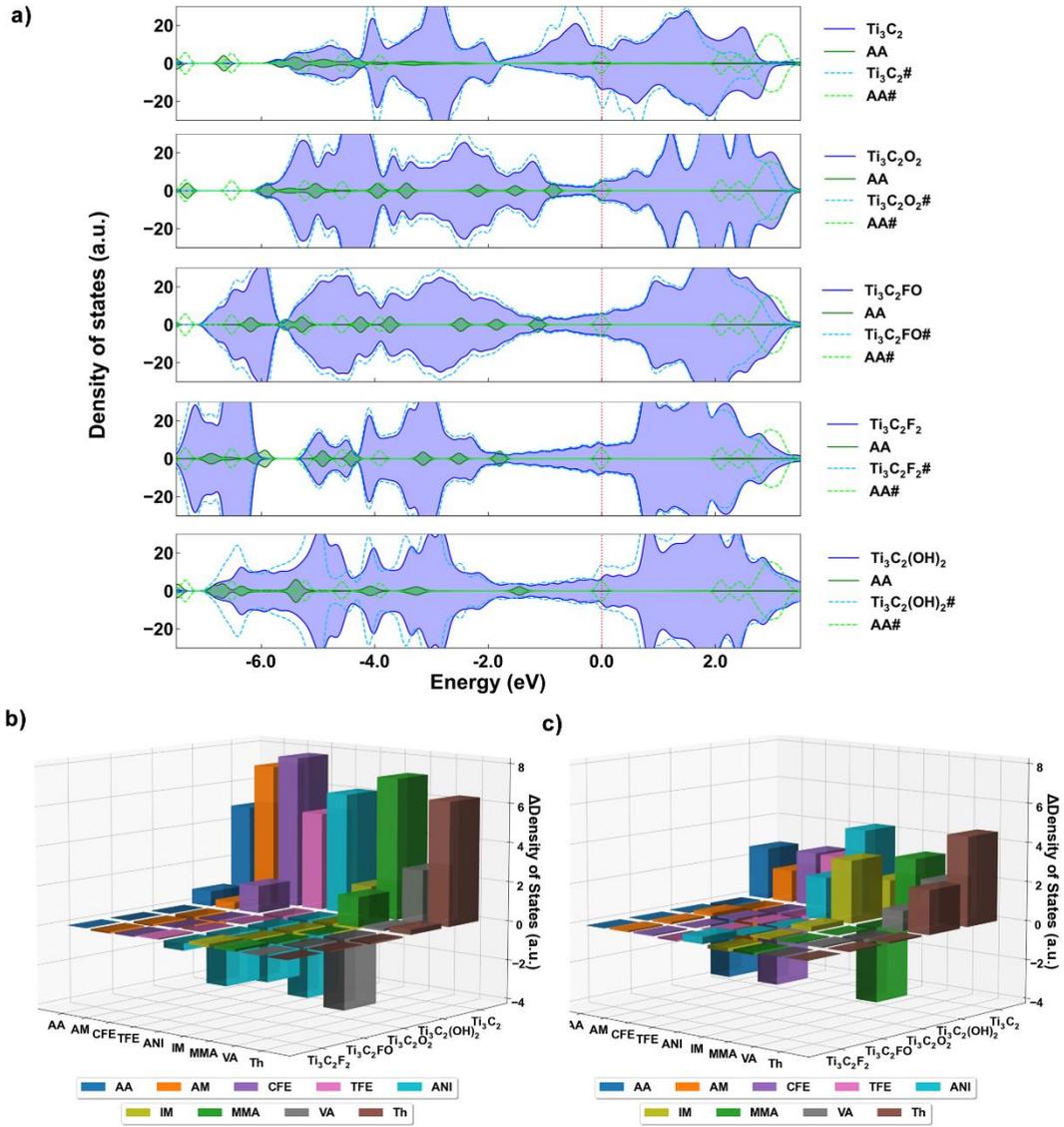

**Figure 5.** a) LDOS of monolayer MXene/AA (solid lines) and the total DOS (dashed lines) of bare MXene and isolated monomer as marked by #. Variation of metallicity of MXene due to polymer adsorption as reflected by the change of b) spin up, and c) spin down decomposed LDOS of MXene/Monomer at the Fermi energy level $E_f$.

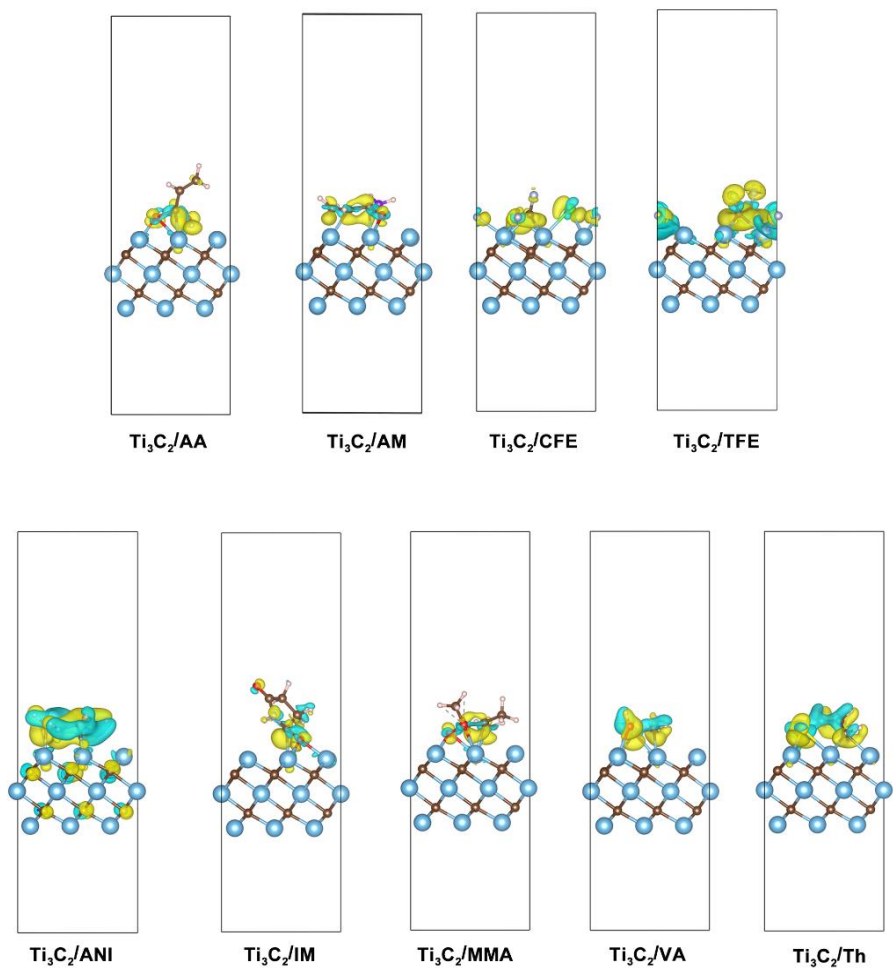

**Figure 6.** Charge transfer across the interlayer with isosurface value 0.01.

To find out the redistribution of charges associated with the adsorption of monomers, we also analyze the charge transfer between the monomers and $Ti_3C_2T_2$. The isosurface plot for the differential charge transfer plot of $Ti_3C_2$/Monomer is shown in **Figure 6**. Regarding the bare $Ti_3C_2$/Molecule composite, the migration of electrons is all from the MXene layer to the monomers, which originated from the relatively strong interaction with forming covalent bonds. For quantifying the intensity of charge redistribution, the amount of electron migration is calculated as shown in **Figure 7a**.

Taking the $Ti_3C_2$/AA structure as an example, the isosurface of differential charge density (DCD) $\Delta\rho(\mathbf{r})$ and its line-profile curves $\Delta\rho(z)$ along z-direction by integrating the in-plane DCD among x-y plane. The total amount of transferred electrons along z-direction $\Delta q(z)$ is calculated by the integration of the line-profile curves $\Delta\rho(z)$ from the bottom expressed as $\Delta q(z) = \int_{-\infty}^{z} \Delta\rho(z')\, dz'$.[42]

For those monomers decomposed above $Ti_3C_2$, the redistribution of electrons is apparent, around 0.76 e migrated from MXene to CFE, 0.61 e to the TFE, and 0.60 e to the Th (**Figure 7b**). We find that the charge transfer occurs at functional groups like O or N for monomers if the carbon chain is the backbone of the molecule. On the contrary, for carbocyclic monomers like ANI, IM, and Th containing planar aromatic groups, the charge transfer occurs in the whole molecule. This can be explained by the breaking of the $\pi$ bond induced by the adsorption process. Another reason is the relatively planar structure associated with the carbocyclic monomer. A large reaction surface ensured by the flatter monomers provides more active sites for the exfoliation process, resulting in the enhanced fixation of the monomers.

Different from the planar monomers, for those molecules containing carbon chain, functional groups like oxygen dominated -COOH, -COOR, -CONH$_2$, and -OH are the active groups for binding with the surface Ti atoms of $Ti_3C_2$. Similarly, compared with the monolayer structures, bilayer $Ti_3C_2$ also exhibits the same tendency but transfers slightly more electrons from MXene to monomers in most cases.

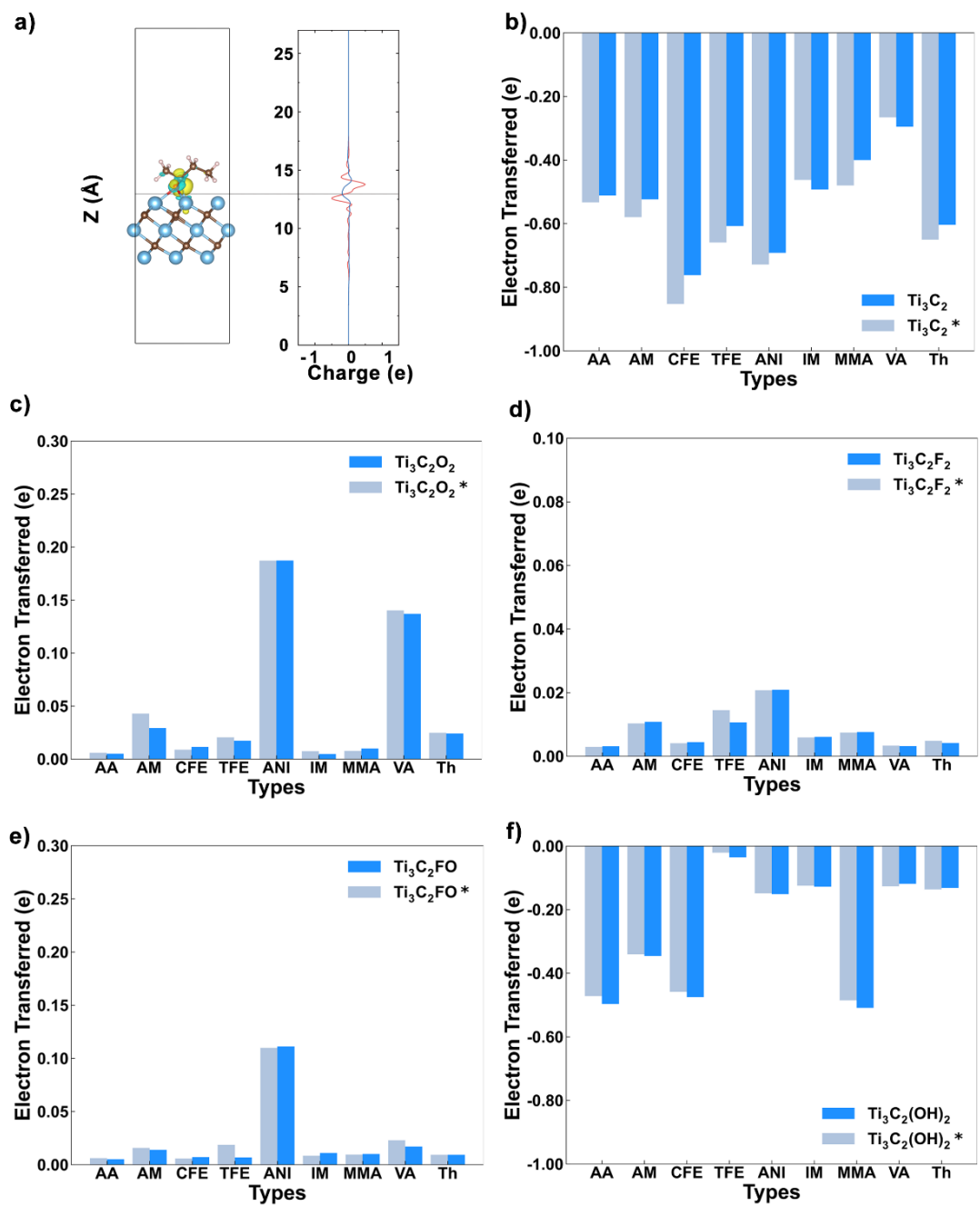

**Figure 7.** Schematic diagram of DCD. a) An example DCD of $Ti_3C_2$/AA. The right panel shows the amount of transferred charge $\Delta q$ and plane-averaged differential charge density $\Delta \rho$ (along $z$ direction) as represented by blue and red, respectively. The left panel is the isosurface plots with blue (yellow) color denoting the loss (accumulation) of electrons. b), c), d), e), and f) Transferred charge across both the monolayer and bilayer $Ti_3C_2$/Monomer, $Ti_3C_2O_2$/Monomer, $Ti_3C_2F_2$/Monomer, $Ti_3C_2FO$/Monomer, and $Ti_3C_2(OH)_2$/Monomer, respectively.

For OH- functionalized MXene/monomer composites, similar migration of the electrons is found during the adsorption process (**Figure 7f**). Due to the abundant surface H atoms on MXene, the binding behavior is especially strengthened with those monomers who have -COOH, -COOR, and -CONH$_2$ functional groups, which lead to 0.35 ~ 0.73 e transferred. Noticeably, for CFE, the decomposition initiates with the breaking of C-Cl bond and the formation of the C-H bond, which also leads to a relatively large electron migration of about 0.48 e. Strikingly different from the pristine $Ti_3C_2$, the trend of charge transfer reverses in the cases of $Ti_3C_2O_2$, $Ti_3C_2FO$, and $Ti_3C_2F_2$, the electrons are transferred from polymer monomers to the oxygen and fluorine functionalized $Ti_3C_2T_2$. Since the surfaces of $Ti_3C_2O_2$, $Ti_3C_2FO$ and $Ti_3C_2F_2$ are fully compensated and relatively inert, as shown in **Figures 7c, 7d,** and **7e**, the amount of charge transfer is significantly suppressed (no more than 0.2 e per each monomer). Nevertheless, for polymers with long chains in experiments, the amount of charge donation from the polymer could still be significant owing to the highly-dense functional groups. For $Ti_3C_2O_2$/Monomer composites, the transferred charges are higher than F terminated or F, O mixed terminated composites because the not fully filled p orbitals of oxygens tend to form weak vdW bonding with H, especially for the planer ANI with 7 H atoms (0.19 e). For the $Ti_3C_2F_2$/Monomer case, the fluorine atoms are tightly bonded to the Ti atoms on the surface of MXene and no more electrons are available for extra covalent bonds, and the charge transfer of $Ti_3C_2F_2$/Molecule is negligible with the highest amounts of 0.021 e. For the $Ti_3C_2FO$/Monomer case, as expected, the electron migration is weakened due to the replacement of half O atoms with F atoms. Therefore, with functional groups of MXene, the formation of the MXene/Molecule is hindered due to weak vdW interaction between the monomers and the $Ti_3C_2T_2$. As shown in **Figure 7**, in both cases, the 5$^{th}$ molecule ANI has the strongest ability to donate electrons to MXene, 0.19 e, 0.11 e, and 0.021 e for $Ti_3C_2O_2$, $Ti_3C_2FO$, and $Ti_3C_2F_2$, respectively. Polymers with such monomers would be able to disperse MXene while simultaneously enhancing the metallicity of MXene. The different amount of charge transfer per unit functional group suggests that mixing MXene with a designed polymer with designed functional monomers would be able to modulate the carrier density in MXene, which can further tune the properties such as conductivity, plasma frequency, dielectric properties, etc.

## CONCLUSION

In summary, through first-principles calculations, we have theoretically explored the exfoliation/adsorption behavior of MXene with nine different monomers of PAA, PAM, PCFE, PTFE, PANI, PI, PMMA, PVA, and PTh, which are already fabricated experimentally. We find the naked $Ti_3C_2$ tends to have a strong interaction with halogen- and oxygen-contained polymers with the formation of covalent bonds. For the $Ti_3C_2O_2$, $Ti_3C_2FO$, and $Ti_3C_2F_2$, the monomers form vdW interactions and charge transfer with the MXene occurs, while for $Ti_3C_2(OH)_2$, a similar tendency to naked $Ti_3C_2$ is found. Our work proves the strong modulated ability of exfoliation with monomers containing different chemical groups. Polymers containing planar functional groups like thiophene leads to promoted interaction compared with chain structure. In particular, we have demonstrated that monomers with halogen atoms (F and Cl) or carbon cyclic with the planar structure are feasible for the fabrication of MXene-based composites with a compact structure. For the AA, CFE, TFE, MMA, and Th monomers, there are molecular levels in the proximity of the Fermi level, and these close-lying states may be sampled under finite voltage which will act as resonant channels for conduction and thermally activated electronic emission. This will be very promising for using conjugated chain molecules where the positions of levels of frontier orbitals can be modulated with chain length. Our work suggests polymers could be used for chemically exfoliating MXene and functionalization to modulate the conductivity and dielectric polarization.

## ASSOCIATED CONTENT

**Supporting Information**

Initial adsorption structures, optimized structures for calculation, and density of states (DOS) of both monolayer and bilayer MXene/Monomer (DOC)

## AUTHOR INFORMATION

**Corresponding Author**

**Yongqing Cai** − Joint Key Laboratory of the Ministry of Education, Institute of Applied Physics and Materials Engineering, University of Macau, Taipa, Macau 999078, P.R. China; orcid.org/0000-0002-3565-574X; Email: yongqingcai@um.edu.mo

**Authors**

**Qiye Guan** − Joint Key Laboratory of the Ministry of Education, Institute of Applied Physics and Materials Engineering, University of Macau, Taipa, Macau 999078, P.R. China

**Hejin Yan** − Joint Key Laboratory of the Ministry of Education, Institute of Applied Physics and Materials Engineering, University of Macau, Taipa, Macau 999078, P.R. China

**Notes**

The authors declare no competing financial interest.

ACKNOWLEDGMENTS

The authors acknowledge the Natural Science Foundation of China (Grant 22022309), and the Natural Science Foundation of Guangdong Province, China (2021A1515010024), the University of Macau (SRG2019-00179-IAPME) and the Science and Technology Development Fund from Macau SAR (FDCT-0163/2019/A3). This work was performed in part at the High-Performance Computing Cluster (HPCC) which is supported by the Information and Communication Technology Office (ICTO) of the University of Macau.

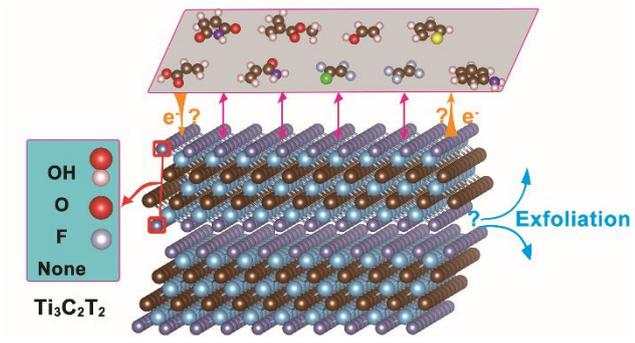

TOC Graphic